# Business models for the simulation hypothesis

**Evangelos Katsamakas**

**Gabelli School of Business, Fordham University, New York, NY, USA**

First version: 5/20/23; This version: 4/6/24

**Abstract**

The simulation hypothesis suggests that we live in a computer simulation. That notion has attracted significant scholarly and popular interest. This article explores the simulation hypothesis from a business perspective. Due to the lack of a name for a universe consistent with the simulation hypothesis, we propose the term *simuverse*. We argue that if we live in a simulation, there must be a business justification. Therefore, we ask: If we live in a simuverse, what is its business model? We identify and explore business model scenarios, such as simuverse as a project, service, or platform. We also explore business model pathways and risk management issues. The article contributes to the simulation hypothesis literature and is the first to provide a business model perspective on the simulation hypothesis. The article discusses theoretical and practical implications and identifies opportunities for future research related to sustainability, digital transformation, and Artificial Intelligence (AI).

## 1. Introduction

The simulation hypothesis suggests that we live in a computer simulation [1]. Popularized by several press articles [2–8], the idea also gave rise to extensive academic literature exploring related foundations and implications. However, the current scholarly literature lacks a business perspective on the simulation hypothesis.

This article proposes a business perspective to the simulation hypothesis. It introduces the *simuverse* term to characterize a universe consistent with the simulation hypothesis. It explores the business justification of the simulation hypothesis and business implications. The main question it asks is: If we live in a simuverse, what is its business model? Why would someone, individual or organization, want to create a simuverse? We propose a novel business model perspective on the simuverse. The perspective focuses on business models for the simuverse and allows for refining the simulation argument.

Ideas related to the simulation hypothesis appear in writings of philosophers like Zhuangzi, Plato, and Descartes [9], movies like *The Matrix*, and essays by computing visionaries like Hans Moravec [10]. Elon Musk and other technology and science thought leaders believe there is a very high probability that we live in a simulation [11]. Early computing pioneers speculated that "the universe might be nothing but a giant computer continually executing formal rules to compute its own evolution" [12].

The primary contribution of the article is that it provides a business perspective on the simulation hypothesis, focusing on the business and economic incentives to create a simuverse. It connects the simulation hypothesis literature with business model concepts. Methodologically, this article takes a systems and futures thinking approach [13–16] and outlines narrative scenarios about the simuverse. The article identifies business models, business model pathways, and risk management issues. We discuss theoretical and practical implications, simulation as a tool for understanding our world, and future directions for sustainability, digital transformation, and AI. Overall, the article and the future directions seek to advance a novel research agenda on the business and economics of the simuverse.

The following section provides a background on the simulation hypothesis and discusses computation in the simuverse. Section 3 develops the article's main perspective, exploring business model scenarios, pathways, and risk management, and recognizes the value of simulation as a method. Section 4 provides theoretical and practical implications of our business model perspective on the simulation hypothesis. We conclude in section 5.

## 2. Background

This section reviews the related literature, clarifies essential concepts, and provides a foundation for the main ideas proposed in section 3.

## 2.1 The simulation argument

The simulation hypothesis derives from the simulation argument [1] following probabilistic reasoning. The simulation argument assumes substrate independence, meaning a biological brain is not required to emulate a human mind and consciousness. It also considers the computing resources needed to run ancestor simulations and the likely vast computing resources available to a post-human civilization. Those considerations lead to the conclusion that advanced post-human civilizations will have sufficient computing power to run "hugely many ancestor simulations even while using only a tiny fraction of their resources for that purpose" (p. 248). Then, the main idea connects three parts: the probability that a civilization at our stage of development will become a post-human civilization, the probability that a post-human civilization will be running ancestor simulations, and the probability that we live in a simulation. If human civilizations do not become extinct, and they run many ancestor simulations, then there is a high chance we live in a simulation. Alternatively, if we do not live in a simulation, our civilization will likely go extinct or not be running simulations. It is also possible that simulated civilizations become post-human civilizations that run their simulations; then, we end up with multiple, but not necessarily infinite, levels of simulations.

## 2.2 Exploring the simulation hypothesis

The simulation hypothesis has implications for the nature of reality and our universe. It gave rise to extensive literature with contributions from philosophers, physicists, and computer scientists. Some of that literature seeks to clarify [17,18], update [19], or extend the simulation argument [20–23]. Other work tests the simulation theory [24] or devises algorithms to support the simulation [25]. We cannot prove we are not in a simulation because any supporting evidence can be simulated [9]; some work formally estimates that the probability of living in a simulation is a little less than 50 percent [26]. Other authors take a more critical view of the simulation argument [27–30] and the simulation hypothesis [31,32]. Variations of the simulation hypothesis include the peer-to-peer simulation hypothesis [33].

The simulation hypothesis provides an alternative explanation for the fine-tuning argument [34]. Fine-tuning refers to the universe being fine-tuned for life [35], explained with either a multiverse hypothesis [36–38] or a design argument. Another stream discusses theological implications [39–41]. Other recent writings explore the potential types of simulation [42] and explore constraints using quantum arguments and experiments [43] or numerical simulations from a physics perspective [44,45].

## 2.3 Simuverse and computation

The simuverse is a universe consistent with the simulation hypothesis. Because the simuverse is a computer simulation, it relies on computation by definition. Computation here is synonymous with well-defined information processing; however, formal treatments of the computation and information concepts can be found in the theory of computation [46] and information theory [47–49].

A view of the world centered on information and computation is gaining a broader acceptance. Information, matter, and energy are the three primary constructs of the universe [50]. From physics to biology and economics, we increasingly understand our world as information and information processing or computation, and we use computational tools to model, explain, and predict the behavior of said systems [51–56]. Cellular automata can model the universe as a computation [57,58], and extensive research explores computation in biology [59–64], cognitive psychology and cognition [65], and living systems as diverse as our brain, the immune system, and ant colonies [66–68].

Some physicists further propose the 'it from bit' concept. In particular: "Otherwise stated, every physical quantity, every it, derives its ultimate significance from bits, binary yes-or-no indications, a conclusion which we epitomize in the phrase, *it from bit.*" (p.309) and that "all things physical are information-theoretic in origin and this is a participatory universe." (p.311) [51]. According to that view, information is physical and represents the ultimate nature of reality so that "the universe easily can be interpreted as a vast simulation." (p.473) [69]. According to other physicists, "maybe at rock bottom, the universe is about information and information processing, and it is matter that emerges as a secondary concept" [70]. The view that everything is computation comes in several variations and sometimes is referred to as pan-computationalism [71–74]. According to related works, the universe is a quantum computer [53,75,76], a mathematical structure likely defined by computable functions [77,78], a holographic universe [79], or an autodidactic universe that learns its laws [80].

The importance of computation and information in our world does not prove the simulation hypothesis but provides evidence in line with the simuverse. Nevertheless, a comprehensive synthesis of the concepts of computation, information, and the simuverse should be the subject of future research, possibly aided by quantum computation [81,82]. In the next section, we focus on a business perspective on the simulation hypothesis.

## 3. A business model perspective on the simuverse

The main question motivating this article is: Why would someone, even in the context of an advanced civilization, want to create a simuverse? We argue that there must be a business justification; therefore, we must consider the business model. While Bostrom (2003) mentions "relatively wealthy individuals who desire to run ancestor-simulations and are free to do so" (p. 255), the motivation and incentives for running a simulation are mainly unexplored. First, "wealthy individuals" may run a simulation, but even for them, there must be an economic or business incentive. We explore relevant scenarios in what follows.

We believe exploring the business incentives for creating a simuverse is essential for comprehensively understanding the simulation hypothesis. The evolution of human civilization has shown that no matter how many resources are available at a particular point of development, humans seek to use those resources as productively as possible. That objective is crucial for further development. Thus, even an advanced civilization will want to use its resources as productively as possible as long as it seeks further growth. It is unlikely that advanced civilization has reached a peak that makes further development impossible.

The simuverse design will depend on the technology of the advanced civilization, but the laws of nature set the hard constraint. Other potential constraints will be intellectual property, legal or regulatory. In the following, we discuss three scenarios related to the business justification of the simulation. All scenarios assume an advanced civilization since running such a simulation is infeasible for any civilization at our stage of development.

### 3.1 Scenario 1: Simuverse as a project

In this scenario, an organization (private or governmental) designs and runs the simuverse as an internal project. The team that launched the simulation must provide a business justification for the project. The project needs to have a purpose and solve a well-defined problem that the organization faces. The purpose of creating the simuverse is possibly related to education (e.g., teaching or learning about the universe), entertainment and media (e.g., providing an experience to target users), research (e.g., historical research, or creating data to train machine learning algorithms), arts and culture (e.g., an exhibit in a museum of the advanced civilization). The actual purpose could be any specific purpose falling into those domains. Moreover, the list of domains is not exhaustive, and many other potential domains can be added.

A simulation involves modeling a system driven by the need to solve a problem articulated as a set of well-defined questions [83]. Those questions help define the boundary of the system captured in the simulation. Moreover, the organization will optimally use available resources. The organization will optimize the simulation running time to get the simulation results.

In addition, the organization will want the project to have an impact. The organization will want to derive actionable insights that will lead to policy interventions for system improvement. Those improvements could be incremental or transformational. The expected impact to justify the simuverse project would depend on the magnitude of dedicated resources.

Lastly, considering those issues can provide additional insights into the simulation hypothesis. For instance, assuming scenario 1, a simulation of the universe running for fifteen billion years may suggest that this project does not make productive use of time, and this might be evidence against the simulation hypothesis.

## 3.2 Scenario 2: Simuverse as a service (Saas)

Many of the above concerns apply in this scenario, but now the simuverse provider serves external clients, necessitating a business model. A business model is a blueprint of how a company does business, and it defines the logic of the firm. We can assume a business model ontology with two elements: how a company creates value and how it captures value through a revenue mechanism [84–87]. The provider of the Saas serves a market (clients) willing to pay to access the Saas. The service could be related to entertainment and media, education, research, or any other domain, as discussed in 3.1. The revenue mechanism could be a one-time fee, a subscription fee, or a combination. The provider could offer multiple versions that could appeal to different client groups. A freemium model is also possible.

The Saas business model has several benefits for the clients. They do not need to invest in building their own simuverse. They can pay for what they need as a utility service on-demand. At the same time, the Saas provider can achieve economies of scale, optimize security and operational efficiency, and collect data about how clients use the Saas. In turn, the provider can exploit the collected data to optimize the client experience and the efficiency of operations.

## 3.3 Scenario 3: Simuverse as a platform (Saap)

In this scenario, the provider builds and grows a platform business. A platform defines an architecture of participation that facilitates the interaction of multiple groups of participants. One side is the users that value access to the platform. The other side is complementors that build on the platform, offering third-party content or services that add value. The platform's purpose could be related to entertainment and media, education, research, or any other domain or a mix of domains. The revenue mechanism could be users, complementors, or both paying a one-time participation or subscription fee that may depend on platform usage.

A platform business must solve a chicken-and-egg problem, achieve critical mass, and optimize pricing, openness, investment, governance, platform competition, and complementor coopetition [88–93]. Platforms can benefit from network effects [94,95] and artificial intelligence (AI) feedback loops [96,97]. In addition to its business significance [98], the platform concept aligns with Wheeler's notion of a participatory universe [51].

## 3.4 Business model pathways

The former three scenarios could be a progression over time. Namely, the simuverse could start initially as an internal project. However, it can transition to a Saas after its value proposition becomes clear. Still, when the Saas provider identifies opportunities to become a platform, it becomes a Saap.

In conjunction with scenario 2 or 3, the simuverse provider can monetize the data collected by selling them to other third parties on a subscription or one-time basis. That is viable if third parties lack alternative access to such valuable data. Third parties may use the data to build other simulations or provide additional services.

If there are multiple simuverse providers, we must consider business model competition in the simuverse market. It is not required that providers will have identical business models. On the contrary, they will likely differentiate, and multiple business models may co-exist and remain viable in the market. For instance, a Saas provider may compete against a Saap provider. However, network effects, AI feedback loops, supply-side economies of scale, and other factors creating reinforcing feedback loops may lead to a winner-takes-all market.

## 3.5 Risk management

A business perspective on the simulation hypothesis must account for the associated risks. We take into account two types of risk. There are risks for the simuverse operator (simulator) and the agents (humans) within the simuverse.

Humans face several novel challenges and risks in a simuverse. Decision-makers should optimize decisions over all possible worlds [99]. Moreover, "everybody would have to consider the possibility that their actions will be rewarded or punished, perhaps using moral criteria, by their simulators" (Bostrom, 2003, p. 254). Living in a simulation or not, humans are agents seeking to understand how the world works and decipher the rules of the game. They strive to use that understanding to maximize their benefit (welfare) while avoiding catastrophic and existential risks [100,101].

From a risk perspective, the simuverse may suddenly and unexpectedly break down. There are several termination risks [102]. All software systems have defects (bugs), so a software-defined universe could be terminated due to a simulation software bug or a defect of the hardware (substrate) running the simulation. Moreover, a simulation could be terminated by the simulation operator or some other external agent (hacker). For instance, the simuverse operator may be dissatisfied with the simulation performance or results. They can terminate the simulation any moment they like and forget about it or redesign and restart it using different rules. Alternatively, they could use the resources to run a completely different simulation. Alternatively, an advanced outside agent hacks into the simulation, steals resources, spreads viruses, or attempts to take over the resources to run a wholly different simulation. In extreme scenarios, humans aware of the simulation may seek to escape through various methods [11].

In summary, the most significant risk for the simulator is project failure under scenario 1 or business model failure under scenarios 2 and 3. Moreover, this is not intended to be a comprehensive list of scenarios but a first attempt that could stimulate further future exploration.

## 3.6 Simulation as a method

An integral element of our business perspective on the simuverse is the intensive use of simulation as a research method. An early example of a mechanical simulation of our solar system is the Antikythera mechanism that could track and predict the locations of the sun, moon, and all five planets known in antiquity [103]. Simulations today rely on computational modeling using software and run on computer hardware (substratum).

Simulation allows us to explore what could happen, what could have happened, and what other possible worlds and alternative realities could look like. We simulate alternative realities to learn from the process. A simulation is a rigorous approach to doing thought experiments. In addition, thought experiments are closely connected with narrative, storytelling, and literary fiction [104].

The value of simulation is recognized across several disciplines [105], such as the philosophy of science [106–115], sciences and engineering [116], education [117–119], medical training [120], artificial life [121,122], social sciences [123,124], economics and finance [125–127] and information systems [128–131]. Simulation is crucial to understanding space-time in quantum gravity physics [132] and other astrophysical phenomena [133]. Simulation combined with AI can accelerate discovery and scientific progress [134–140]. For instance, Nvidia's whole earth simulation project on climate change mentions that "simulation is the answer" [141]. Lastly, simulation and other modeling methods should strive to reduce opacity [142], making the modeler's invisible hand visible. Recent AI advances, such as Large language models (LLMs) promise innovative applications of simulation research [143,144].

One valuable side-effect of the simulation hypothesis debates is bringing attention to the simulation concept. That provides an opportunity for a broader adoption of simulation as a research method, generating insights for incremental or transformational interventions. A business perspective on the simuverse relies on simulation to make progress.

Figure 1 summarizes the perspective.

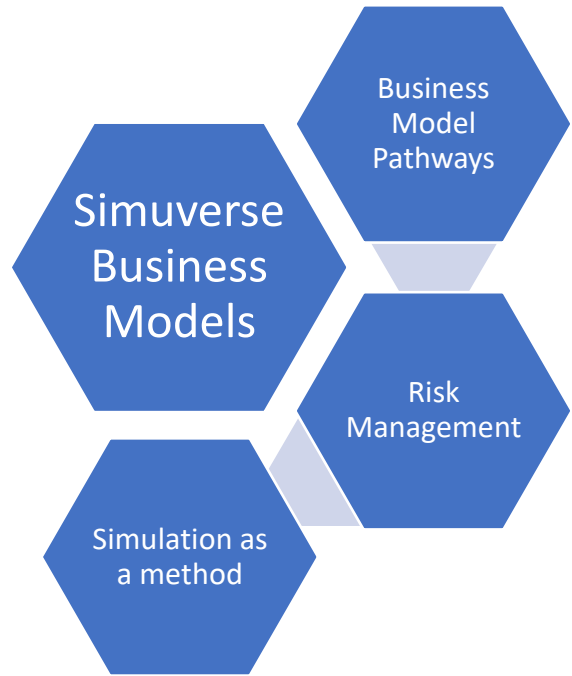

**Figure 1.** Elements of the business model perspective on the simuverse. Current business model scenarios: Simuverse as a project, Simuverse as a service, Simuverse as a platform.

## 4. Discussion

This article introduces the simuverse concept and proposes a novel business perspective on the simulation hypothesis. Because our perspective focuses on simuverse business models, we call it a business model perspective on the simuverse. It includes business model scenarios, pathways, and risk management issues. Overall, the perspective is a step towards a comprehensive understanding of simuverse creation by a sufficiently advanced civilization. We discuss implications for theory and practice.

### 4.1 Business model contributions

Our business model perspective on the simuverse shows that the business model concept can be integrated into the simulation hypothesis literature. Extending business model thinking into the simuverse provides new insights and creates an opportunity for novel business model research that looks very far into the future. Our exploration suggests that creating business models for the distant future would expand both the time horizons of business model research and its design space and could lead to precious innovation. However, it would require developing novel research skills to design qualitative and quantitative thought experiments that appreciate and benefit from complexity.

Research into business models for the distant future would have additional beneficial implications. If we can shed some light on the very distant future, then we can see and appreciate the present more clearly. Moreover, we can pay more attention and accelerate the right pathways toward desirable (eutopic) futures.

A crucial insight of our work is that devising an effective simuverse business model increases the likelihood of the simuverse. That means that if we already live in a simuverse, an organization designed a business model that made the simuverse feasible. Alternatively, suppose we do not live in a simuverse. In that case, an advanced civilization might be able to create a simuverse in the future, provided it can envision and design a suitable simuverse business model.

Our discussion is also relevant for today's organizations. Designing an organization's business model is a crucial activity of senior management because that design affects the survival and prosperity of the organization. This article provides a framework for managers to start thinking about designing business models for the simuverse. Of course, this is a very long-term endeavor. However, our business model perspective on the simuverse can shape the thinking of future managers for the longer term. Our perspective does not offer a comprehensive list of business models. It should be seen as a starting point for more managerial ideation and experimentation, which could be supported by simulation as a tool.

## 4.2 A refinement of the simulation argument

Another significant research contribution of this work is that it expands the simulation hypothesis literature by adding a novel business model perspective on the simuverse. Our analysis has implications for the simulation argument. In particular, our study offers a refinement of the simulation argument, as described next.

Recall that the simulation argument is structured into three propositions (parts). Our analysis pertains to the second part: "The fraction of post-human civilizations that are interested in running a significant number of ancestor simulations is extremely small" (p.54) [19].

According to our analysis, the probability that an advanced civilization is interested in running simulations, especially a large number, increases when a business justification and a plausible, viable business model exist. Under those conditions, the second proposition of the simulation argument is likely false, which makes the simulation hypothesis likely true, provided that civilizations do not go extinct before they become advanced enough to run simulations.

Therefore, our perspective offers a refinement of the simulation argument. That refinement allows a better understanding of the likelihood that the simulation hypothesis is true because it takes into account business and economic incentives. In other words, business and economic incentives are crucial for the creation of the universe.

**4.3 Simuverse and sustainability: envisioning a sustainable simuverse**

One way to think about the simuverse is as a limiting future case with implications for the present. In other words, if we can create a simulation of the universe in the distant future, we need to think very hard today about the responsibility of doing so. A crucial question is: If we could create a simuverse, what is our vision for that? Our research envisions and calls for a sustainable simuverse that merges the simuverse opportunity with the sustainability objectives [145–148]. A comprehensive definition of such a vision is a whole new research stream, but we propose that sustainability should be a design feature. In addition, we must consider relevant risks using those discussed in section 3 as a starting point. Lastly, it is valuable to integrate the meaning, conditions, and implications of post-scarcity into our analysis [151–153].

Those observations also call for a broadening of sustainability's scope and time window. First, sustainability must encapsulate the cosmos, consisting of all the current and future physical and virtual worlds. This suggests that sustainability research and practice needs to expand the system boundaries and think more holistically and dynamically, leveraging complexity sciences to address complex challenges innovatively [15,149,150]. Moreover, it is helpful to expand the time window that frames the sustainability discussions, thinking far back in the past and very long term into the future, considering desirable eutopic pathways. Those directions offer multiple opportunities for future research.

**4.4 Implications for digital transformation and AI**

Our simuverse exploration suggests a rethinking of our digital transformation agenda that takes into account the distant future. In that agenda, digital technologies are primary enablers of sustainable pathways [154]. That requires careful thought experiments (or simulations) about using technology for prosperity because prosperity cannot be taken for granted [155,156]. Complex systems thinking and future-proof design strategies [157] are needed when designing complex digital systems and immersive virtual worlds [158–160]. A complementary technophilosophy approach that addresses core questions of reality, knowledge, and value [9,161]—for instance, what is a good life in a virtual world—is also helpful. Such a systemic perspective on immersive virtual worlds and alignment with human values is necessary if we want to avoid today's social media and AI dysfunctions.

In the longer term, the simuverse could be a fork arising from the evolution of immersive virtual worlds or progressively larger-scale and more sophisticated digital twins [162–166]. Therefore, in the short term, we should focus on designing responsible virtual worlds. As AI progresses toward artificial general intelligence, future virtual worlds could be created by AI or via a collaboration of humans and AI in which the AI acts as a copilot, as in software development today [167]. Alternatively, suppose that the simuverse is created by an artificial superintelligent agent [10,168–170]; this poses additional challenges because the motives and objectives of such an agent, not to speak of a population of agents that compete or cooperate, are difficult to comprehend today fully. Those futures make resolving AI ethics and alignment issues even more crucial today.

Lastly, the direction of technology [156] and future scenarios about AI development and use [171,172] will affect the simuverse scenarios. Therefore, the multiple complex interactions between the simuverse and AI could be another avenue of future research.

## 4.5 Implications for business education

Our research reinforces the significance of a computational and complex systems view of our natural and artificial worlds. Therefore, the opportunity lies in teaching more computational thinking [173], which includes AI, complex systems thinking [13,67,174,175], and simulation as a tool.

The tools we use to understand the world affect crucially the concepts and theories we develop, the way we understand the world, and the actionable insights we derive about what needs to be done to improve the world. Changing the tools may change what we see and decide and what policies we implement to solve problems. We can use simulation to better the world because simulation allows us to *visualize and tell a story* about how a system works, learn from rigorous thought experiments, identify leverage points, and build consensus for change [83].

In the cave allegory explored in Plato's *The Republic*, written around 375 BC, humans live in a cave and can only see the shades of the actual world outside the cave; therefore, escaping the cave is a duty as it allows for understanding the real world. In a sense, simulation is a tool to escape Plato's cave and deal with computational irreducibility challenges. Once we see the value of simulation, its broader adoption in research and education is an obligation.

# 5. Conclusions

The simulation hypothesis is derived from the simulation argument and has attracted significant scholarly and public attention. This article contributes a business perspective to the simulation hypothesis literature. First, we introduce the concept of the *simuverse*, a universe consistent with the simulation hypothesis; the literature lacks such a concept, and it could facilitate the development of the related scholarship. Second, we ask: Why would someone, individual or organization, want to create a simuverse? We argue that there must be a business justification to do so. Therefore, we explore business model scenarios, pathways, and risks.

We show that a business perspective on the simuverse helps refine the simulation argument. Our framework could be used by managers who seek to design business models for the simuverse, and it has implications for sustainability, digital transformation, and AI. Our work reinforces a computational view of the cosmos and suggests an opportunity to intensify the adoption of simulation as a research method.

Suppose we see the simulation argument as a thought experiment. Then, this article extends the thought experiment by adding business model considerations. It provides a first attempt to explore business models for the simuverse. To advance that novel research stream, we identify several opportunities for future research. We also encourage using simulation as a research tool in studying the business and economics of the simuverse.